\documentclass[prd,aps,twocolumn,english,reprint,10pt]{revtex4-2}
\usepackage[T1]{fontenc}
\usepackage{inputenc}
\usepackage{color}
\usepackage{babel}
\usepackage{natbib}
\usepackage{hyperref}
\hypersetup{colorlinks=true, citecolor=blue, urlcolor=blue, linkcolor=magenta}
\usepackage{graphicx}
\usepackage{dcolumn}
\usepackage{array}
\usepackage{amsmath,amssymb}
\usepackage{mathrsfs}
\usepackage{physics}

\begin{document}
\title{Renormalization in Wavelet basis}
\author{Mrinmoy Basak}
\email[Electronic Address: ]{p20180411@goa.bits-pilani.ac.in}
\author{Raghunath Ratabole}
\email[Electronic Address: ]{ratabole@goa.bits-pilani.ac.in}
\affiliation{Deparment of Physics, BITS-Pilani KK Birla Goa Campus,\\
Zuarinagar 403726, Goa, India}
\begin{abstract}
Discrete wavelet-based methods promise to emerge as an excellent framework for the non-perturbative analysis of quantum field theories. In this work, we investigate aspects of renormalization in theories analyzed using wavelet-based methods. We demonstrate the non-perturbative approach of regularization, renormalization, and the emergence of flowing coupling constant within the context of these methods. This is tested on a model of the particle in an attractive Dirac delta function potential in two spatial dimensions, which is known to demonstrate quintessential features found in a typical relativistic quantum field theory.
\end{abstract}

\maketitle

\section{Introduction}
 
The wavelet-based methods have emerged as a strong contender for the non-perturbative analysis of quantum field theories (QFTs). Discrete wavelet-based methods provide ways to analyze these QFTs akin to the Euclidean lattice approach yet allow the study of real-time dynamics \cite{kessler2003wavelet,polyzou2018multi,2017PhDT........99M,Federbush_1995,best2000wavelet,
evenbly2016entanglement,tomboulis2021wavelet,george2022entanglement,https://doi.org/10.48550/arxiv.2207.08294,best2000wavelet}. The discrete and multiscale nature of compactly supported wavelets holds the promise of providing a systematic framework for classical and quantum simulations of these continuum QFTs \cite{9402012,HALLIDAY1995414,Ismail_2003_2,PhysRevA.92.032315,Polyzou:2021cdj}. Discrete wavelets have also been used for analyzing statistical field theories \cite{Polyzou_2013,PhysRevD.87.116011,PhysRevD.95.094501,neuberger2018wavelets}. Methods developed based on continuous wavelets provide a complementary perspective \cite{altaisky2007wavelet,albeverio2009remark,Altaisky_2018,altaisky2013continuous}. Both approaches, discrete and continuous, have considered regularization, renormalization, and gauge invariance in field theories. Wavelet-based representation of light-front quantum field theories has been formulated \cite{Altaisky:2012lda,altaisky2016quantization,PhysRevD.101.096004} to gain an advantage from the unique properties of being on the light-front \cite{https://doi.org/10.48550/arxiv.2002.04981}. In this paper, we investigate the approach to renormalization when working within the framework of Daubechies wavelet-based quantum theories.

The fundamental theories of elementary particles and their interactions
are described by local QFTs formulated on 3+1 dimensional Minkowski
space-time. The quantization of these field theories is usually done
using canonical quantization approach or via path integral methods. As
a part of the quantization process, it is common to resolve the field
using the plane wave basis into its momentum modes. In the plane wave
basis, the free field part of the Hamiltonian represents these momentum
modes as uncoupled oscillators, while the interaction Hamiltonian
represents the couplings between the different momentum modes of the field. When computing the S-matrix elements, it is common
to adopt a manifestly covariant approach to perturbation theory to
effectively deal with the ultraviolet divergences and re-express the
theory in terms of renormalized masses and couplings \cite{Peskin:1995ev,Mandl:1985bg}. However, in
doing so, the central role played by the Hamiltonian eigenvalue problem 
does get compromised. 

The lattice approach allows one to analyze QFTs beyond perturbation
theory systematically. One defines the field on a Euclidean lattice
with a presumed underlying lattice cutoff. The quantum field
theory is studied as an equivalent statistical field theory with the
partition function defined as a path integral over the Euclidean action.
The discrete nature of the lattice makes the field theory computationally
tractable. The presence of the explicit cutoff regulates the ultraviolet
divergences (beyond perturbation theory), but it explicitly violates
Euclidean invariance. There is often a trade-off between the need
for non-perturbative analysis and the desire for full covariance.
Within the lattice approach, the continuum limit of QFT is obtained
by maintaining criticality in the limit of vanishing cutoff. With
the exception of the Hamiltonian lattice approach, Euclidean lattice
makes it challenging to work with the Hamiltonian energy eigenvalue
problem directly. 

The discrete wavelet-based formulation of quantum field theory allows
one to commit to the Hamiltonian framework while maintaining the discreteness
of the lattice approach, yet not compromise the continuum nature of
space. Daubechies wavelets and scaling functions constitute an orthonormal
basis (generally referred to as wavelet basis in the rest of the paper) of compactly supported functions \cite{doi:10.1137/1.9781611970104,chui1992introduction,daubechies1988orthonormal}. Roughly speaking, each basis
function is characterized by its location (translation index) and
length scale (resolution). The quantum fields, when expanded in the wavelet
basis, lead to its representation as an infinite
sequence of operators characterized by a location and resolution index.
This approach allows natural volume and resolution truncations of
the QFT. The truncated theory is an ordinary quantum mechanical theory
with multiple discrete degrees of freedom organized by location and
length scale. The maximum resolution plays the role of ultraviolet
cutoff. 

In this paper, we use an example of the two-dimensional Dirac delta
functional potential to illustrate aspects of renormalization within
the discrete wavelet-based approach. Several authors \cite{Atkinson:1975vv,Gosdzinsky:1990vz,Jackiw:1991je,doi:10.1119/1.16675,doi:10.1063/1.531271,doi:10.1063/1.532350,Cavalcanti:1998jx,doi:10.1119/1.19051,CAMBLONG200114,Nyeo2000RegularizationMF,geltman2011bound,PhysRevA.65.052123,https://doi.org/10.48550/arxiv.cond-mat/0305631,PhysRevA.88.064101,Erman_2017,ANWONG20182547,Loran_2022,Pazarbas__2019} have studied
this potential to understand the nuances of renormalization in an
elementary setting. In this work, we showcase the emergence of asymptotic freedom
within the wavelet approach. The outline of the paper is as follows:
In Section II, we introduce the essential basic properties of Daubechies
wavelets, following which, in Section III, we show how this discrete
wavelet basis can be used to address problems in quantum mechanics.
In Section IV, after a brief discussion on renormalization within
the wavelet framework, we present the analysis of the
two-dimensional Dirac delta function potential using wavelet based approach. Section VI contains concluding remarks and future outlook. 

\section{Daubechies Wavelets}
In this section, we summarize the key aspects of the construction of Daubechies wavelet basis and their associated properties, which are used through out this paper \cite{doi:10.1137/1.9781611970104,chui1992introduction,PhysRevD.95.094501,PhysRevD.87.116011,kessler2003wavelet,daubechies1988orthonormal}. The basis elements consist of scaling functions and wavelet functions, which are generated starting from a single function $s(x)$ (often referred to as the mother scaling function) defined through the linear renormalization group equation:
\begin{eqnarray}
s(x)&=& \sum_{n=0}^{2K-1} h_n \hat{D} \hat{T}^n s(x)\label{eq:scaling}
\end{eqnarray}
$\hat{D}$ and $\hat{T}$ denote the scaling and translation operations respectively. These unitary operations are defined by
\begin{eqnarray}
\hat{D}s(x)=\sqrt{2}s(2x),\qquad \hat{T}s(x)=s(x-1)\label{eq:D and T}
\end{eqnarray}
$\hat{T}$ translates the function as a whole by one unit to the right without altering its form, while $\hat{D}$ shrinks the support of the function by a factor of two while maintaining its norm,
\begin{eqnarray}
\int s(x)dx=1 \label{eq:norm}
\end{eqnarray}
In eq (\ref{eq:scaling}), $K$ represents a fixed integer that in turn will determine the extent of smoothness and the support of the basis functions.
 
\begin{figure}[ht]
\includegraphics[scale=0.525]{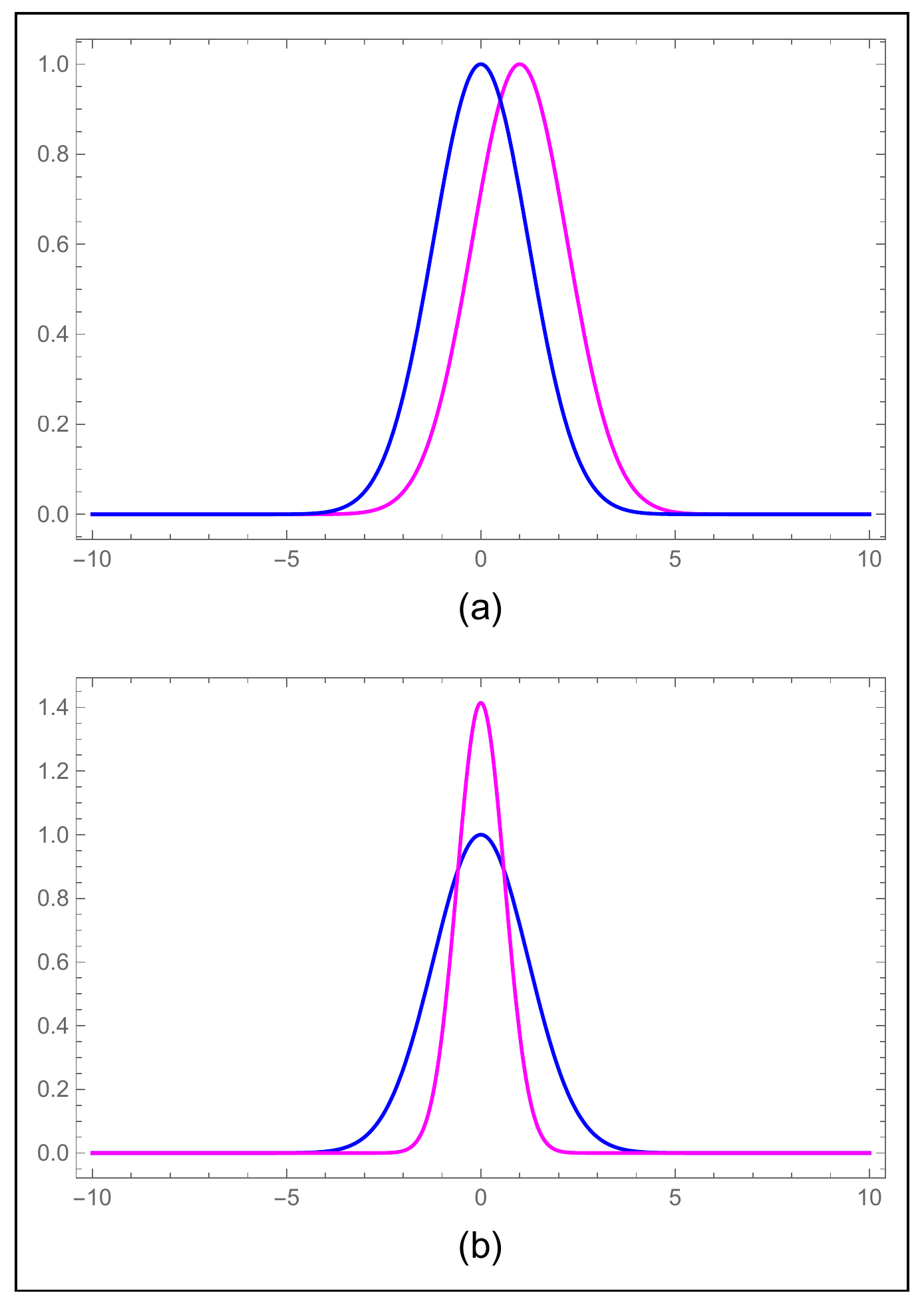}
\caption{\label{fig1}Any generic function $f(x)$ (blue color). The action of (a) Translation operator $\hat{T}$ and (b) Scaling operator $\hat{D}$ on that function(magenta color).}
\end{figure}

The action of the operators $\hat{D}$ and $\hat{T}$ are shown for a typical function $f(x)$ in fig (\ref{fig1}).

\begin{figure}[ht]
\includegraphics[scale=0.3]{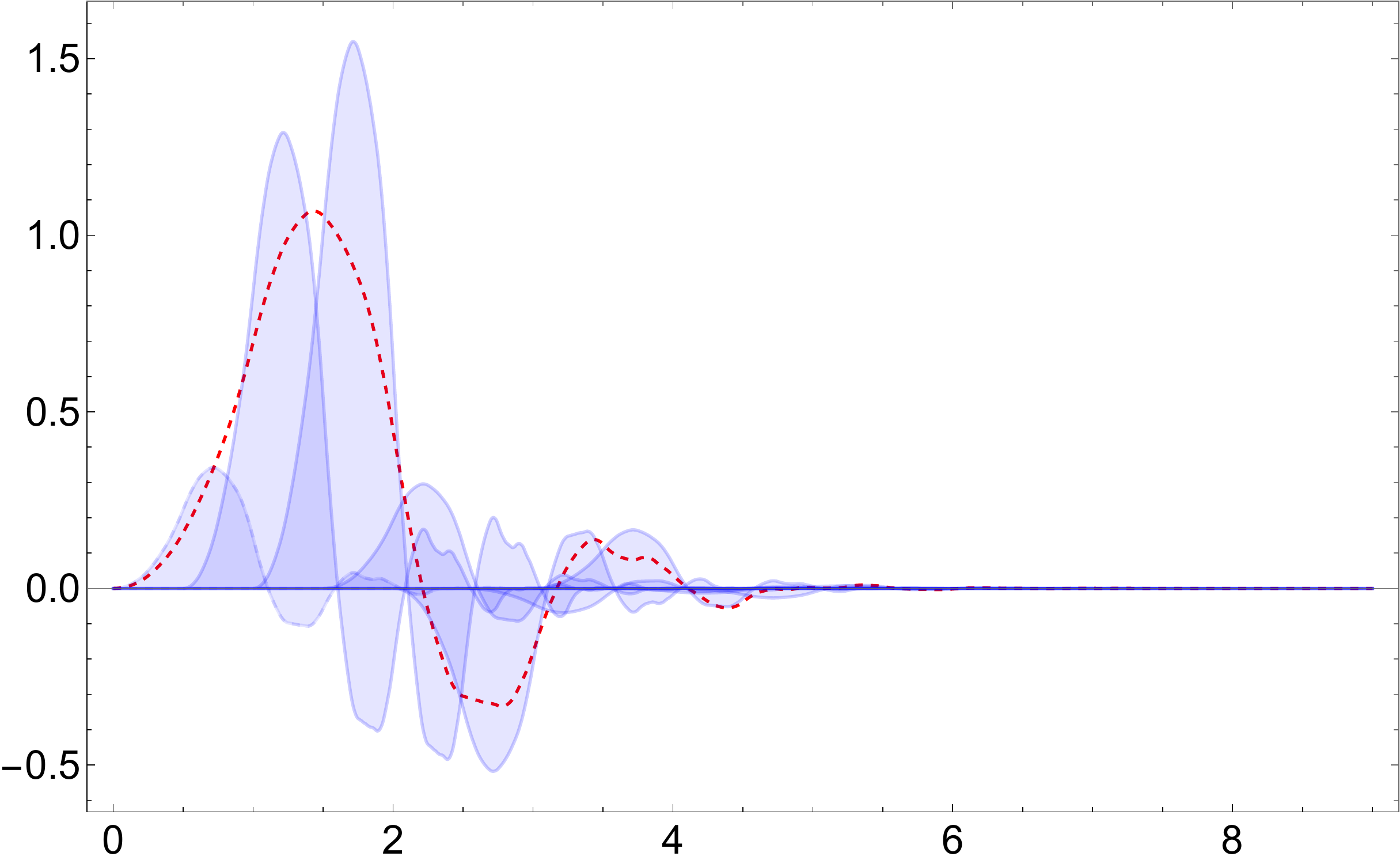}
\caption{\label{fig2}The red dotted line shows the mother scaling function $s(x)$ for $K=5$ formed as a weighted sum of $10$ translated copies of $s(x)$ scaled to half of the original support.}
\end{figure}

Eq (\ref{eq:scaling}) defines $s(x)$ as a specific linear combination of $2K$ translated and scaled copies of itself. This is visually represented in fig (\ref{fig2}) for $K=5$.

Given with the solution of eq (\ref{eq:scaling}) we define the $k$-th resolution scaling functions by applying $n$ unit translations followed by $k$ dyadic scale transformations on the mother scaling function,
\begin{eqnarray}
s^k_n(x):=\hat{D}^k\hat{T}^n s(x)
\end{eqnarray}
The scaling functions $s^k_n(x)$ are orthonormal,
\begin{eqnarray}
\label{eq:scaling_scaling_ortho}
\int s^k_m(x)s^k_n(x)dx&=&\delta_{mn}
\end{eqnarray}
and arbitrary linear combination of these functions generate the space $\mathscr{H}^k$ of resolution $k$:
\begin{eqnarray}
\label{eq:f(x)_from_k_scaling_functions}
\mathscr{H}^k=\{f(x)|f(x)=\sum_{-\infty}^{\infty}f_n s_n^k(x),|f_n|^2<\infty\}
\end{eqnarray}

From scaling equation eq (\ref{eq:scaling}) and eq (\ref{eq:f(x)_from_k_scaling_functions}), it follows that,
\begin{eqnarray}
\mathscr{H}^{k}\subset \mathscr{H}^{k+1}
\end{eqnarray}
and more generally, for any $m>0$
\begin{eqnarray}
\mathscr{H}^{k}\subset \mathscr{H}^{k+m}
\end{eqnarray}
this means that the $k$-th resolution space is a linear subspaces of the $(k+m)$-th resolution space. 

\begin{figure}[ht]
\includegraphics[scale=.525]{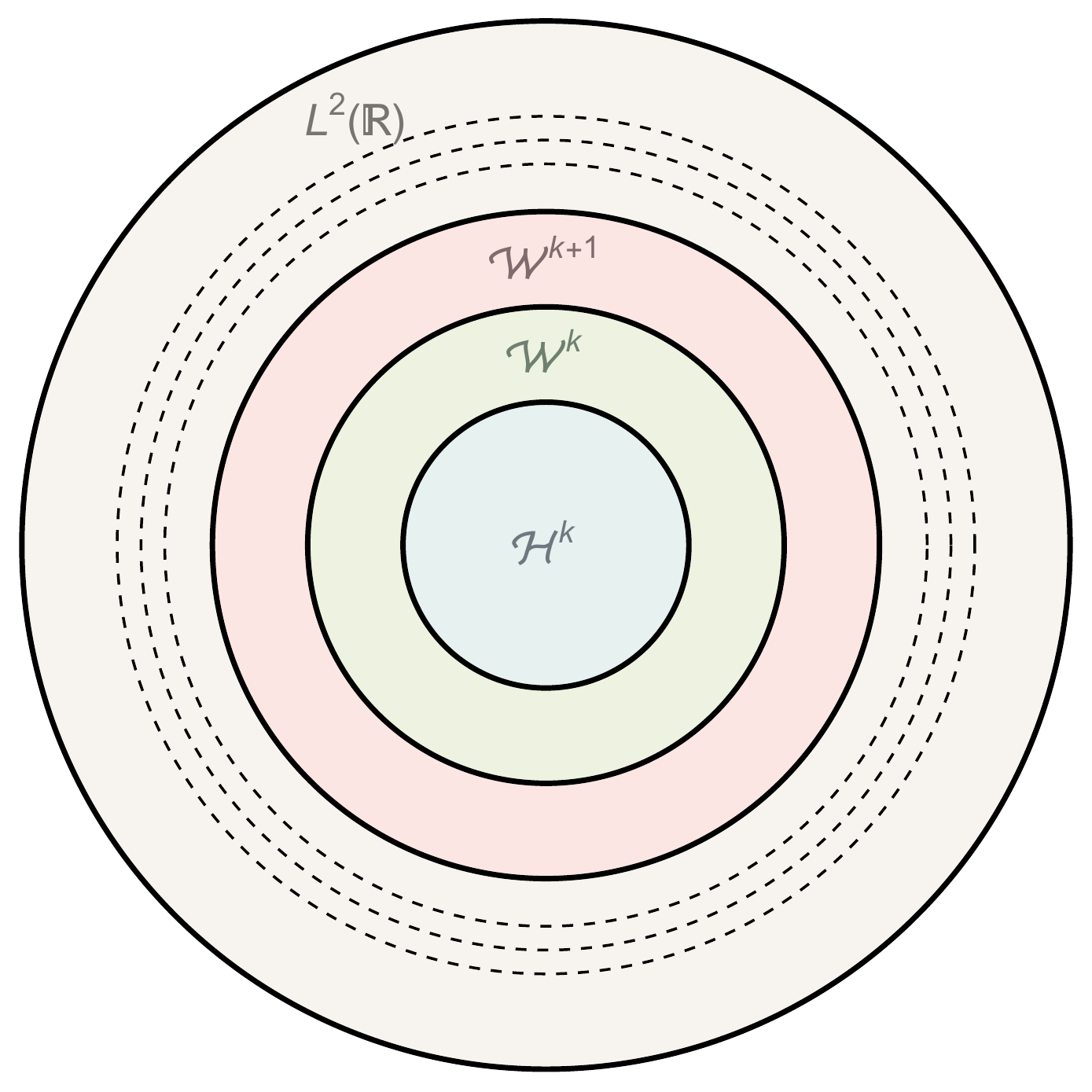}
\caption{\label{fig4}Euler diagram for spanning of Hilbert space with wavelet basis.}
\end{figure}

Now we define the mother wavelet function $w(x)$ such that it is orthogonal to the mother scaling function:
\begin{eqnarray}
\label{eq:wavelet}
w(x)=\sum_{n=0}^{2K-1}g_n \hat{D}\hat{T}^n s(x)
\end{eqnarray}
where
\begin{eqnarray}
g_n=(-1)^n h_{2K-1-n}
\end{eqnarray}
The mother wavelet function $w(x)$ will be used to construct the orthogonal complement $w^k(x)$ of $\mathscr{H}^k$ in $\mathscr{H}^{k+1}$. Towards this end, we define the wavelet function $w_n^k(x)$ as:
\begin{eqnarray}
w^k_n(x):=\hat{D}^k\hat{T}^n w(x)
\end{eqnarray}
The wavelet functions are orthonormal 
\begin{eqnarray}
\label{eq:wavelet_wavelet_ortho}
\int w^k_m(x)w^{l}_n(x)dx&=&\delta_{mn}\delta_{kl}
\end{eqnarray}
and arbitrary linear combination of these generates the space $\mathscr{W}^{k}$ of resolution $k$:

\begin{eqnarray}
\label{eq:wavelet_expansion}
\mathscr{W}^k=\{f(x)|f(x)=\sum_{-\infty}^{\infty}f_n w_n^k(x),|f_n|^2<\infty\}
\end{eqnarray}
By design, the scaling functions and wavelet functions are orthogonal to each other 
\begin{eqnarray}
\label{eq:scaling_wavelet_ortho}
\int s^k_m(x)w^{k+l}_n(x)dx&=&0\quad,\quad l\geq 0
\end{eqnarray}
and 
\begin{eqnarray}
\label{eq:res_rel}
\mathscr{H}^{k+1}=\mathscr{H}^k\oplus \mathscr{W}^k
\end{eqnarray}
The space of square integrable real functions, $\mathsf{L}^2(\mathbb{R})$, can be generated by recursive use of eq (\ref{eq:res_rel}).
\begin{eqnarray}
\label{eq:hilbert_space}
\mathsf{L}^2(\mathbb{R})=\mathscr{H}^k\oplus\mathscr{W}^{k}\oplus \mathscr{W}^{k+1}\oplus \mathscr{W}^{k+2}...
\end{eqnarray}
This has been visually represented in fig (\ref{fig4}).

There are two possible choices of basis for $\mathscr{H}^k$, one could either work with the scaling function basis of resolution $k$ ($\{s^{k}_n(x)\}_{n=-\infty}^{\infty}$) or a combination of resolution $k-1$ scaling and wavelet functions ($\{s^{k-1}_n(x)\}_{n=-\infty}^{\infty}\cup \{w^{k-1}_n(x)\}_{n=-\infty}^{\infty}$). These two basis are related to each other through an orthogonal transformation given by:
\begin{eqnarray}
s^{k-1}_n(x)=\sum_{l=0}^{2K-1}h_l s^{k}_{2n+l}(x)\quad\quad\quad\quad \\
w^{k-1}_n(x)=\sum_{l=0}^{2K-1}g_l s^{k}_{2n+l}(x)\quad\quad\quad\quad \\
s^{k}_{n}(x)=\sum_{m} h_{n-2m}s^{k-1}_m(x)+\sum_{m}g_{n-2m}w^{k-1}_m(x)
\end{eqnarray}

By induction, for any fixed value of $k$, the scaling functions and wavelet functions  
\begin{eqnarray}
\{s^k_n\}^{\infty}_{n=-\infty}\cup \{w^m_n\}^{\infty,\infty}_{n=-\infty,m=k}
\end{eqnarray}
will form the basis for $\mathsf{L}^2(\mathbb{R})$. Any square integrable function, $f(x)$ can be expanded in this basis
\begin{eqnarray}
\label{eq:square_integrable_decomposition}
f(x)=\sum_{n=-\infty}^{\infty}f^s_n s^k_n(x)+\sum_{n=-\infty}^\infty \sum_{l=k}^{\infty}f^{w,l}_{n}w^l_n(x)
\end{eqnarray}
such that 
\begin{eqnarray}
\label{eq:square_integrable_condition}
\sum_{n=-\infty}^{\infty}|f^s_n|^2+\sum_{n=-\infty}^{\infty}\sum_{l=k}^{\infty}|f^{w,l}_{n}|^2<\infty
\end{eqnarray}

An alternative way to construct $\mathsf{L}^2(\mathbb{R})$ is via an infinite resolution limit of $\mathscr{H}^k$.
\begin{eqnarray}
\mathsf{L}^2(\mathbb{R})=\lim_{k \to \infty} \mathscr{H}^k
\end{eqnarray}
We are going to use this approach for the rest of the paper.

The real weights $h_n$ associated with the order $K$ mother scaling function can be determined for an integer value of $K$ by solving the system of equations eq (\ref{eq:hn}), eq (\ref{eq:normality}) and eq (\ref{eq:Orthogonal_to_(k-1)}).
\begin{eqnarray}
\label{eq:hn}
\sum_{n=0}^{2K-1}h_n =\sqrt{2} \quad\quad\quad\quad \\
\label{eq:normality}
\sum_{n=0}^{2K-1}h_nh_{n-2m}=\delta_{m0} \quad\quad\quad \\
\label{eq:Orthogonal_to_(k-1)}
\nonumber \sum_{n=0}^{2K-1}n^mg_n=\sum_{n=0}^{2K-1}n^m(-1)^n h_{2K-1-n}\\ 
=0,\quad m<K \quad\quad\quad
\end{eqnarray}
Eq (\ref{eq:hn}) is the necessary condition for the scaling equation to have a solution. Eq (\ref{eq:normality}) tells us that the integer translation of scaling functions are orthonormal to each other. Eq (\ref{eq:Orthogonal_to_(k-1)}) ensures that the linear combination of integer translates of the wavelet functions are orthogonal to degree $K-1$ polynomials.\cite{kessler2003wavelet}

The coefficients $h_n$ for $K=1$, $2$ and $3$ are given in the table (\ref{tab:h_n_coefficients}).

\begin{table}[h]
\begin{tabular}{|c|c|c|c|}
\hline
$h_n$ & $K=1$ & $K=2$ & $K=3$\\
\hline
$h_0$ & $1/\sqrt{2}$ & $(1+\sqrt{3}/4\sqrt{2})$ & $(1+\sqrt{10}+\sqrt{5+2\sqrt{10}})/16\sqrt{2}$\\
\hline
$h_1$ & $1/\sqrt{2}$ & $(3+\sqrt{3})/4\sqrt{2}$ & $(5+\sqrt{10}+3\sqrt{5+2\sqrt{10}})/16\sqrt{2}$\\
\hline
$h_2$ & $0$ & $(3-\sqrt{3})/4\sqrt{2}$ & $(10-2\sqrt{10}+2\sqrt{5+2\sqrt{10}})/16\sqrt{2}$\\
\hline
$h_3$ & $0$ & $(1-\sqrt{3})/4\sqrt{2}$ & $(10-2\sqrt{10}-2\sqrt{5+2\sqrt{10}})/16\sqrt{2}$\\
\hline
$h_4$ & $0$ & $0$ & $(5+\sqrt{10}-3\sqrt{5+2\sqrt{10}})/16\sqrt{2}$\\
\hline
$h_5$ & $0$ & $0$ & $(1+\sqrt{10}-\sqrt{5+2\sqrt{10}})/16\sqrt{2}$\\
\hline
\end{tabular}
\caption{\label{tab:h_n_coefficients}h-coefficients of Daubechies wavelets for different values of $K$.}
\end{table}

\begin{figure}[ht]
\includegraphics[scale=.374]{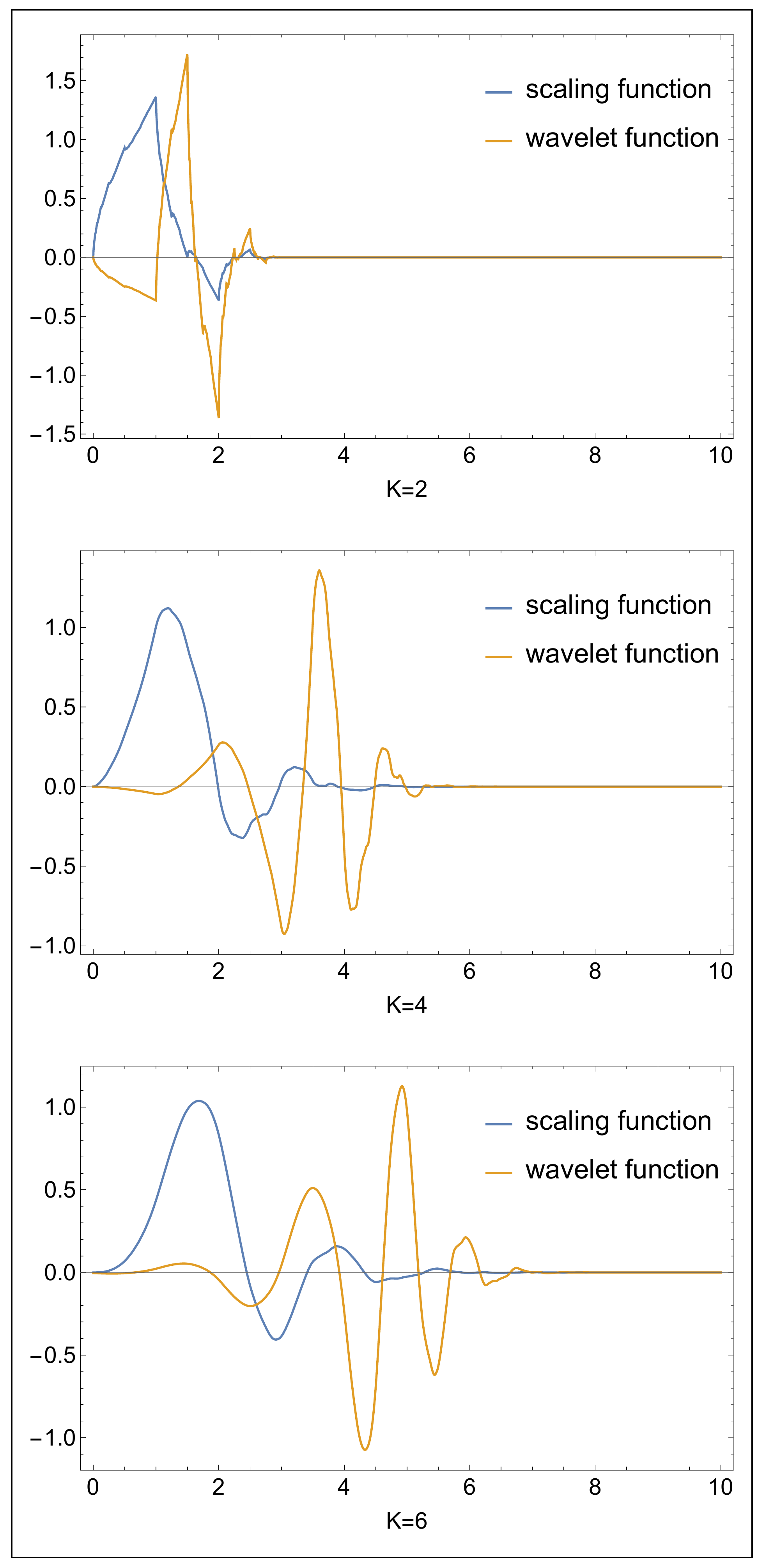}
\caption{\label{fig:incresing_K_scaling_wavelet_functions}Scaling and Wavelet functions for different values of $K$.}
\end{figure}

Using the weights $h_n$ we can determine $s(x)$ and $w(x)$ at each point $x$ from eq (\ref{eq:scaling}) and eq (\ref{eq:wavelet}) \cite{Kessler2002ScatteringCW,article}. It can be shown that the mother scaling function $s(x)$ and mother wavelet function $w(x)$ have compact support on the interval $[0,2K-1]$. A graphical view of $s(x)$ and $w(x)$ for sample value of $K=2$, $4$ and $6$ is shown in fig (\ref{fig:incresing_K_scaling_wavelet_functions}).The basis functions $s^k_n(x)$ and $w^k_n(x)$ have compact support smaller by a factor $2^k$ in comparison with the $s(x)$ and $w(x)$.
\begin{eqnarray}
s^k_n(x),w^k_n(x)\neq 0 &\forall& x\in\left(\frac{(0-n)}{2^k},\frac{(2K-1-n)}{2^k}\right)\nonumber\\
&\implies& \text{support size}=\frac{(2K-1)}{2^k}\nonumber
\end{eqnarray}

The degree of analyticity of the basis functions depends on the value of $K$. For example, The basis functions for $K=1$, $2$ are not differentiable, $K=3$, $4$ are singly differentiable, $K=6$ are doubly differentiable, and so on.

The extension of the basis in two dimensions can be done by forming direct products of one dimensional scaling and wavelet functions.
\begin{eqnarray}
\label{eq:gen_scaling}
s^k_{n_1,n_2}(\textbf{x}):=s^k_{n_1}(x_1)s^k_{n_2}(x_2)
\end{eqnarray}
and we introduce another notation $w^m_{\textbf{n},\alpha}(\textbf{x})$, which we call generalize wavelets having the following forms,
\begin{eqnarray}
\label{eq:gen_wavelet_1}
w^m_{n_1,n_2,1,k_2}(\textbf{x})&:=&s^k_{n_1}(x_1)w^{k_2}_{n_2}(x_2)\\
\label{eq:gen_wavelet_2}
w^m_{n_1,n_2,2,k_1}(\textbf{x})&:=&w^{k_1}_{n_1}(x_1)s^{k}_{n_2}(x_2)\\
\label{eq:gen_wavelet_3}
w^m_{n_1,n_2,3,k_1,k_2}(\textbf{x})&:=&w^{k_1}_{n_1}(x_1)w^{k_2}_{n_2}(x_2)
\end{eqnarray}
$m$ represents the smallest wavelet scale appearing in the product and $\alpha$ indicates the values of $k_1,k_2$ and the four types of products that are appearing in the basis function. Any square integrable function in two dimensions can be expanded in this basis as follows:

\begin{eqnarray}
f(x_1,x_2)=\sum_{n_1,n_2} f^s_{n_1,n_2}s^k_{n_1,n_2}(\textbf{x})+\quad\quad\quad\quad\quad\nonumber\\
\sum_{\substack{n_1,n_2\\ k_2\geq k}} f^{w_1,m}_{n_1,n_2}w^m_{n_1,n_2,1,k_2}(\textbf{x})+\sum_{\substack{n_1,n_2\\ k_1\geq k}} f^{w_2,m}_{n_1,n_2}w^m_{n_1,n_2,2,k_1}(\textbf{x})\nonumber\\
+\sum_{\substack{n_1,n_2\\ k_1,k_2\geq k}} f^{w_3,m}_{n_1,n_2}w^m_{n_1,n_2,3,k_1,k_2}(\textbf{x})\quad\quad\quad\quad\quad
\end{eqnarray}
such that
\begin{eqnarray}
\sum_{n_1,n_2} |f^s_{n_1,n_2}|^2+\sum_{\substack{n_1,n_2\\ k_2\geq k}} |f^{w_1,m}_{n_1,n_2}|^2+\sum_{\substack{n_1,n_2\\ k_1\geq k}} |f^{w_2,m}_{n_1,n_2}|^2\nonumber\\
+\sum_{\substack{n_1,n_2\\ k_1,k_2\geq k}}|f^{w_3,m}_{n_1,n_2}|^2\leq \infty\quad\quad\quad\quad\quad
\end{eqnarray}
Here, the summation over $n_1$, $n_2$ goes from $-\infty$ to $\infty$ and summation over $k_1$, $k_2$ goes from $k$ to $\infty$.
\section{2d dirac delta function Potential problem in wavelet basis}
The energy eigenvalue equation for the 2D-Dirac delta function potential in natural units ($m=1,\hbar=1$) is given by,
\begin{eqnarray}
\label{eq:DDF_eigen_value}
\left[-\frac{1}{2}\sum_{i=1}^{2}\frac{\partial^2}{\partial x_i^2}-g\delta(x_1)\delta(x_2)\right]\psi(x_1,x_2)=E\psi(x_1,x_2)\quad
\end{eqnarray}
We approximate the state space of the system to the resolution subspace $\mathscr{H}^k$. Within this approximation, by expanding eigenfunction $\psi(x_1,x_2)$ in the scaling function basis,
\begin{eqnarray}
\psi(x_1,x_2)=\sum_{n_1,n_2}\psi^{k}_{n_1,n_2}s^k_{n_1,n_2}(\textbf{x})
\end{eqnarray} 
we can express eq (\ref{eq:DDF_eigen_value}) as a matrix eigenvalue equation
\begin{eqnarray}
\sum_{n_3,n_4}H^k_{s,n_1,n_2:n_3,n_4}\psi^k_{n_3,n_4}=E\psi^k_{n_1,n_2}
\end{eqnarray}
where, the Hamiltonian matrix elements are given by,
\begin{eqnarray}
H_{s,n_1,n_2:n_3,n_4}^{k}= \int \Bigg(-\frac{s^k_{n_1,n_2}(x_1,x_2)}{2}\sum_{i=1}^{2}\frac{\partial^2 s^k_{n_3,n_4}(x_1,x_2)}{\partial x_i^2}\nonumber\\
-g s^k_{n_1,n_2}(x_1,x_2)\delta(x_1)\delta(x_2)s^k_{n_3,n_4}(x_1,x_2)\Bigg)dx_1dx_2\quad\quad
\label{eq:Hamiltonian_matrix}
\end{eqnarray}
Using integration by parts and the compact support of the scaling functions, we can rewrite eq (\ref{eq:Hamiltonian_matrix}) as:
\begin{eqnarray}
H_{s,n_1,n_2:n_3,n_4}^{k}= \int \Bigg(\frac{1}{2}\sum_{i=1}^{2}\frac{\partial s^k_{n_1,n_2}(x_1,x_2)}{\partial x_i}\frac{\partial s^k_{n_3,n_4}(x_1,x_2)}{\partial x_i}\nonumber\\
-g s^k_{n_1,n_2}(x_1,x_2)\delta(x_1)\delta(x_2)s^k_{n_3,n_4}(x_1,x_2)\Bigg)dx_1dx_2\quad\quad
\end{eqnarray}
From the separable nature of the two-dimensional scaling basis functions, we can rewrite the Hamiltonian matrix elements involving different overlap integrals as:
\begin{eqnarray}
H^k_{s,n_1,n_2:n_3,n_4}&=&\frac{1}{2}\left[D^{k}_{ss,n_1n_3}\times \delta_{n_2n_4}+D^{k}_{ss,n_2n_4}\times \delta_{n_1n_3}\right]\nonumber\\
&& + g I^k_{ss,n_1n_3}\times  I^k_{ss,n_2n_4}
\end{eqnarray}
where
\begin{eqnarray}
\label{eq:s_prime_s_prime_1}
D^k_{ss,mn}&=&\int_{-\infty}^{\infty} s^{k'}_{m}(x)s^{k'}_n(x) dx\\
\label{eq:delta_scaling_scaling_1}
I^{k}_{ss,mn}&=&\int_{-\infty}^{\infty} \delta(x)s^k_{m}(x)s^{k}_n(x)dx
\end{eqnarray}
We can evaluate these integrations analytically from the properties of scaling functions using the procedure described in the appendix\ref{Appen:Overlap_Integrals}.

\begin{figure}[ht]
\includegraphics[scale=.375]{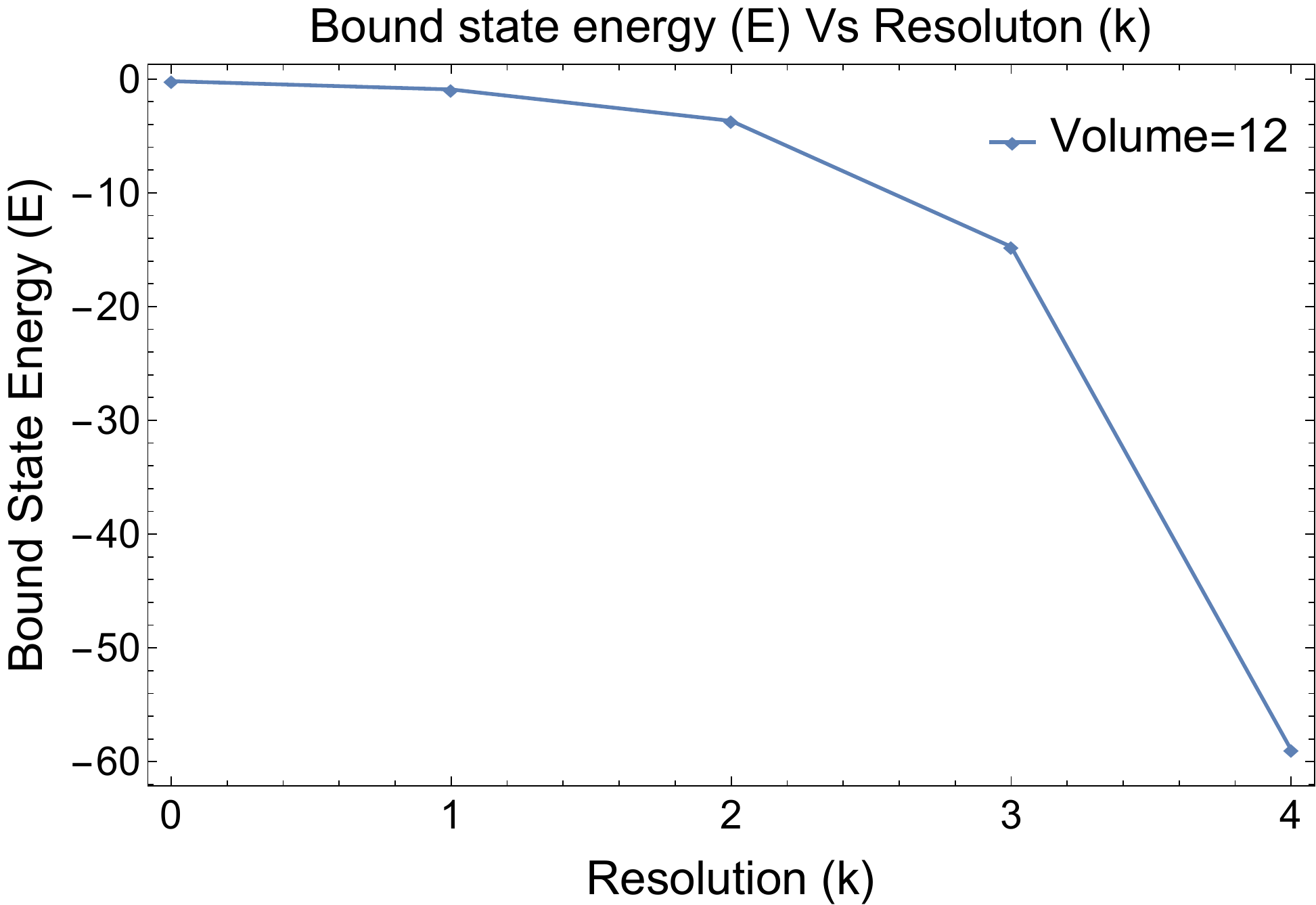}
\caption{\label{fig:divergence_of_bound_state}Negative divergence of bound state energy for fixed volume ($V=12$) with fixed coupling constant ($g=1.848694$) versus increasing resolution plot.}
\end{figure}

The resulting Hamiltonian matrix has to be truncated for carrying out numerical computation of its eigenvalues. The first kind of truncation is volume truncation which identifies the region of physical space accessible to the physical system. In the present problem, we define the volume truncation by $-L\leq x,y\leq L$. The other truncation involves choosing a value for resolution $k$. This implies inclusion of all lengths scales down to $(2K-1)/2^k$ and excluding all lengths scales finer than this limit.  For a fixed value of bare coupling constant $g$ and resolution, the low-lying eigenvalues remain unchanged with increasing $L$. This saturation of eigenvalues is intuitively expected as most of the dynamics happens around the origin due to short range of the potential. From here on, all computational results will be reported by choosing an adequately large value of $L$. There exists a minimum value of coupling constant $g_{min}$, beyond which one gets exactly one bound state. For example, if $L=6$ and resolution $k=0$, this minimum value is around $g_{min}=1.18$. For any fixed value of the bare coupling constant, the bound state eigenvalue diverges to negative infinity ref. fig (\ref{fig:divergence_of_bound_state}).
This signifies the appearance of ultraviolet divergences within the wavelet based framework. Concluding line in the section on renormalization: We deal with these divergences by renormalizing the theory. The following section demonstrates the renormalization and continuum limit within this wavelet-based approach.
\section{Renormalization}
The concept of renormalization first arose in context of relativistic quantum field theories. The QFTs defined in the continuum have points in
the underlying space-time that can come arbitrarily close to each
other. In other words, there does not exist
any short-distance (ultraviolet) cutoff. When these quantum field theories are analyzed perturbatively,
one encounters divergences at each order of perturbation theory, whose
origin can be traced to the lack of underlying short-distance cutoff.
The concept of renormalization provides the essential element to derive
physical predictions from the perturbative QFT. At
each order of perturbation theory, the ultraviolet divergences are
regulated by the introduction of an artificial ultraviolet cutoff, following
which the dependence of the bare couplings on the ultraviolet cutoff
is determined by demanding that it reproduces the experimental values
of a finite set of physical observables. For a perturbatively renormalizable
theory, this process of renormalization renders all observables of
the quantum field theory finite and ultraviolet cutoff independent.
In other words, a local limit can be established for a renormalizable theory within the perturbative framework. 

Within the wavelet framework, each quantum state of the system can described as an expansion in scaling and wavelet functions. The expansion coefficients of the scaling functions describe contributions from the coarsest length scale down to $(2K-1)/2^k$, while the expansion coefficients of the wavelet functions represent contribution on all lengths scales finer than $(2K-1)/2^k$. We can impose a short distance cutoff to regulate the theory at a non-perturbative level by truncating the basis function expansion to include only the scaling functions. In other words, the Hilbert space of the system is restricted to $\mathscr{H}^k$. Likewise, all operators (for example, the Hamiltonian) are defined in terms of their action on $\mathscr{H}^k$. The bare coupling constants of the truncated theory are tuned to reproduce the experimental values of a finite set of physical observables. The process of renormalization consists of constructing the local limit by solving a series of truncated theories with increasing resolution. 

\begin{figure}[ht]
\includegraphics[scale=.375]{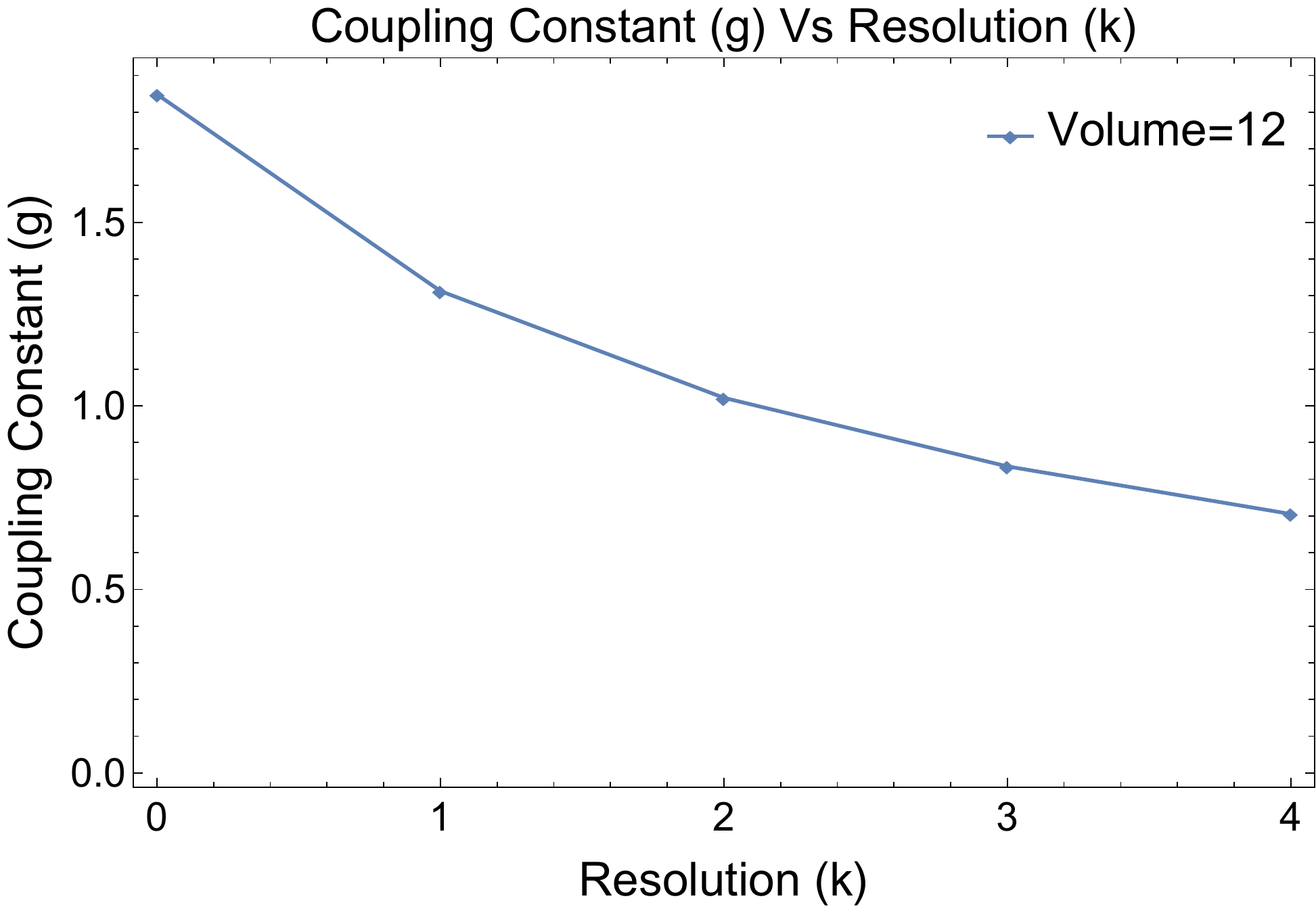}
\caption{\label{fig:resolution_vs_coupling_constant}Renormalized coupling constant $(\tilde{g}(k))$ versus Resolution($k$) plot}
\end{figure}

We showcase the application of this wavelet based approach to renormalization in the context of two dimensional Dirac delta function potential. We have shown in the previous section that, for a fixed value of bare coupling constant, the ground state energy diverges to negative infinity with increasing resolution. In order to have a physically meaningful theory containing a bound state, we bring in a renormalization prescription that the theory truncated at resolution $k$ should reproduce the ground state eigenvalue which we fix at $-1$. We tune the coupling constant value in order for the truncated theory to reproduce this experimental observable. Repeating this process for a series of truncated theories with increasing resolution, we arrive at the observation that the coupling constant flows with resolution. In particular, it becomes weaker with increasing resolution as is expected in the context of this problem. See fig (\ref{fig:resolution_vs_coupling_constant}) and table (\ref{Tab:coupling_constant_resolution}).

\begin{table}[h]
\begin{tabular}{|c|c|}
\hline
Resolution($k$) & Coupling constant$(g)$ \\
\hline
$4$ & $0.7053401$ \\

$3$ & $0.8349675$ \\

$2$ & $1.021796$ \\

$1$ & $1.312652$ \\

$0$ & $1.848694$ \\
\hline
\end{tabular}
\caption{\label{Tab:coupling_constant_resolution} The values of renormalized coupling constant with different resolution cut-off.}
\end{table}

\section{Summary And Conclusions}
The purpose of this work has been to examine renormalization in a discrete wavelet based quantum theory. The attractive two-dimensional Dirac delta function potential was chosen as the model of study as it contains many of the non-trivial features that are observed in a relativistic quantum field theory such as ultraviolet divergences, asymptotic freedom and dimensional transmutation. Working with models such as this one, will provide insights which will valuable when working with realistic quantum field theories within the wavelet based framework.

For quantum systems with finite number of dynamical variables, the operator energy eigenvalue problem is converted to matrix eigenvalue problem using the discrete Daubechies wavelet basis, in which the rows and columns of the matrix can be organized by length scales. The off-diagonal Hamiltonian matrix elements have a natural interpretation of coupling between length scales. Specifically, each Hamiltonian matrix element carries a pair 'location' and resolution indices. By imposing an upper and lower bound on the 'location' index, one can define the region of physical space in which the system would be studied. Setting an upper bound on the resolution index essentially amounts to imposing an ultraviolet cutoff and as such plays the role of the ultraviolet regulator at a non-perturbative level.

We have shown that, if the bare coupling constant is held fixed, then as the resolution is increased the ground state energy diverges as is expected. To make physical sense of this theory, one demands that the bare coupling constant flows with resolution in such a way as to maintain the physical value of the ground state energy. The coupling constant becomes weaker as resolution is increased which attests the asymptotically free nature of the two dimensional Dirac delta function potential. 

In the context of QFTs, the field operator can be expanded in terms of scaling and wavelet basis function with operator valued coefficients. This decomposition leads to the quantum field (the operator valued distribution) being replaced in terms of a countably infinite number of operators with different spatial resolutions. One define a volume truncation by retaining only those basis function terms in the field operator expansion, that have support lying within a specified volume. The resolution truncation admits only those basis function terms in the field operators expansion that are coarser than a specified resolution. This truncated QFT, which is now a theory with finite number of degrees of freedom, should in principle be solvable. The infinite volume and infinite resolution limit needs to be constructed as a limit of a sequence of truncated theories. Further investigations in this direction are highly desirable.  
\section*{acknowledgement}
MB would like to thank BITS Pilani K K Birla Goa Campus for providing the necessary infrastructure and financial support to conduct this work. MB would also like to thank Dr. Rudranil Basu for useful discussions.   
\appendix*
\section{Overlap Integrals}
\label{Appen:Overlap_Integrals}
Here we describe the analytical method to compute the overlap integrals involving product of scaling functions and their derivatives, such as the one appearing in eq (\ref{eq:s_prime_s_prime_1}). This method, due to Beylkin \cite{doi:10.1137/0729097} has been described in \cite{PhysRevD.87.116011,2017PhDT........99M}. We describe this approach in the context of the overlap integrals appeared in eq (\ref{eq:s_prime_s_prime_1}) and eq (\ref{eq:delta_scaling_scaling_1}). The results for the overlap integrals involving the delta function, eq (\ref{eq:delta_scaling_scaling_1}) are new and have not been reported in literature previously.

The following identities are used to compute these integrals:
\begin{eqnarray}
\int s^k_n(x)dx=\frac{1}{\sqrt{2^k}}\\
DT^{2k}=T^kD\\
\label{eq:derivative _translation_commutation}
\frac{d}{dx}D=2D\frac{d}{dx}\\\
Dx=2xD\\
Tx=(x-1)T
\end{eqnarray}
In addition, the scaling equation, the definition of wavelet and the derivatives of these equations are used in the following form:
\begin{eqnarray}
\label{eq:scaling_eq_2}
s^k_m(x)&=&\sum_n H_{mn}s^{k+1}_n(x)\\
\label{eq:scaling_prime_eq_2}
s^{k'}_m(x)&=&2\sum_n H_{mn}s^{k+1'}_n(x)
\end{eqnarray} 
Where
\begin{eqnarray}
H_{mn}=h_{n-2m}
\end{eqnarray}
The general matrix element of the kinetic energy term can be the found out from the following matrix element.
\begin{eqnarray}
\label{eq:s_prime_s_prime}
D^k_{ss,mn}&=&\int s^{k'}_{m}(x)s^{k'}_n(x) dx
\end{eqnarray}
Using eq (\ref{eq:derivative _translation_commutation}), eq (\ref{eq:s_prime_s_prime}) can be expressed in the following form,
\begin{eqnarray}
\label{eq:scale_k_to_0_scaling_scaling}
D^k_{ss,mn}&=&2^{2k}D_{ss,mn}
\end{eqnarray}
where 
\begin{eqnarray}
\label{eq:k_0_s_prime_s_prime}
D_{ss,mn}=\int s'_m(x)s'_n(x)dx
\end{eqnarray}
Using translational invariance, we can rewrite eq (\ref{eq:k_0_s_prime_s_prime}) as,
\begin{eqnarray}
D_{ss,mn}=D_{ss,0(n-m)}=\int s'(x)s'_{n-m}(x)dx
\end{eqnarray}
For $K=3$, the scaling function and their derivatives have support on $[0,5]$. This means $D_{ss,0q}$ will be non-zero only for the values of $q$ lying between $-4$ to $4$. These nine non-trivial integrals are related to each other through a set of equations which we derive below. 

Using the unitary nature of scaling operator $\hat{D}$, we reduce the resolutions of the scaling function by a factor of $-1$:
\begin{eqnarray}
D_{ss,0q}=D^{-1}_{ss,0q}=\int s^{-1'}(x)s^{-1'}_q(x) dx
\end{eqnarray}
Now, raising the resolution by a factor of $1$ using eq (\ref{eq:scaling_prime_eq_2}), we arrive at a set of homogeneous equations for $D_{ss,0q}$.
\begin{eqnarray}
D_{ss,0q}&=&4\sum_{p,r} H_{0p}H_{qr} \int s'_p(x)s'_r(x)dx\nonumber\\
\label{eq:Homogeneous_equation_1}
&=&4\sum_{p,r}h_p h_{r-2q}D_{ss,0(r-p)}
\end{eqnarray}
We now determine an inhomogeneous equation for variable $D_{ss,0q}$ which, in conjunction with the homogeneous set of equations (\ref{eq:Homogeneous_equation_1}), allows us to uniquely determine $D_{ss,0q}$. Our starting point is the decomposition $1$,$x$ and $x^2$ in terms of the scaling functions where $\expval{x^n}$ is called the $nth$ moment of the scaling function. 
\begin{eqnarray}
\label{eq:partition_of_unity_1}
1=&&\sum_n s_n(x)\\
\label{eq:x_in_scaling}
x=&&\sum_n (n+\expval{x})s_{n}(x)
\end{eqnarray}
and
\begin{eqnarray}
\label{eq:x^2_in_scaling}
x^2=\sum_n\left(n^2+2n\expval{x}+\expval{x^2}\right)s_{n}(x)
\end{eqnarray}
Here, $\expval{x^n}=\int x^n s(x)dx$ are called the moments of the scaling function. Eq (\ref{eq:partition_of_unity_1}) is a property of the scaling functions called 'partition of unity'.

Differentiating eq (\ref{eq:x_in_scaling}) and using eq (\ref{eq:partition_of_unity_1}) we get,
\begin{eqnarray}
\label{eq:partition_of_unity_2}
1=\sum_{n}ns'_n(x)
\end{eqnarray}
Yet another differentiation, but this time of eq (\ref{eq:x^2_in_scaling}) followed by use of eq (\ref{eq:partition_of_unity_2}) we get,
\begin{eqnarray}
2x=\sum_n \left(n^2+2n\expval{x}\right)s'_n(x)=\sum_n n^2s'_n(x)+2\expval{x}\nonumber\\
\end{eqnarray}
Multiplying both sides with $s'(x)$ and integrating, we arrive at
\begin{eqnarray}
\int 2xs'(x)dx&=&\sum_n \left(n^2\int s'(x)s'_n(x)dx+2\expval{x}\int s'(x)dx \right)\nonumber\\
\implies -2&=&\sum_n n^2 \int s'(x)s'_n(x)dx\nonumber\\
\label{eq:inhomogeneous_equation_1}
\implies -2&=&\sum_n n^2 D_{ss,0n}
\end{eqnarray}
the inhomogeneous equation eq (\ref{eq:inhomogeneous_equation_1}) for variables $D_{ss,0q}$. 

The linear system of equations eq (\ref{eq:Homogeneous_equation_1}) and eq (\ref{eq:inhomogeneous_equation_1}) can be solved exactly. These values of $D_{ss,0n}$ turn out to be rational numbers and were first calculated in reference \cite{doi:10.1137/0729097}. See table ({\ref{Tab2}}).
\begin{table}[h]
\begin{tabular}{|c|c|}
\hline
Integrals & Values \\
\hline
$D_{ss,0(-4)}$ & $-3/560$ \\

$D_{ss,0(-3)}$ & $-4/35$ \\

$D_{ss,0(-2)}$ & $92/105$ \\

$D_{ss,0(-1)}$ & $-356/105$ \\

$D_{ss,00}$ & $295/56$ \\

$D_{ss,01}$ & $-356/105$\\

$D_{ss,02}$ & $92/105$\\

$D_{ss,03}$ & $-4/35$\\

$D_{ss,04}$ & $-3/560$\\
\hline
\end{tabular}
\caption{\label{Tab2} The values of overlap integrals of product of derivative of scaling functions.}
\end{table}

The overlap integration involving the product of delta function and two scaling function eq (\ref{eq:delta_scaling_scaling_1}) can be expressed in terms of the $0th$ resolution scaling functions using the property of scaling operator $\hat{D}$ and changing the variable of the integration:
\begin{eqnarray}
\label{eq:delta_scaling_scaling_2}
I^{k}_{ss,mn}&=&\int \delta(x)s^k_{m}(x)s^{k}_n(x)dx\nonumber\\
&=&2^{k} I_{ss,mn}
\end{eqnarray}
where,
\begin{eqnarray}
I_{ss,mn}=\int \delta(x) s_m(x)s_n(x)dx
\end{eqnarray}
Now, from the unitary nature of the operator $\hat{D}$ we can express $I_{mn}$ in terms of $I^{-1}_{mn}$ as follows,
\begin{eqnarray}
I^{-1}_{ss,mn}&=&\int \delta(x)s^{-1}_m(x)s^{-1}_n(x)dx\\
&=&\frac{1}{2}\int \delta(x)s_m(x)s_n(x)dx\nonumber\\
\label{eq:delta_scaling_high_low_res}
\implies  I_{ss,mn}&=&2\times I^{-1}_{ss,mn}
\end{eqnarray}
and using eq (\ref{eq:scaling_eq_2}) we can find out the set of homogeneous equation for the integral $I_{mn}$,
\begin{eqnarray}
\label{eq:delta_scaling_homogeneous_equation}
I_{ss,mn}=2\times \sum_{p,q} H_{mp}H_{nq}I_{ss,pq}
\end{eqnarray} 
To get the inhomogeneous equation we start with the definition of delta function and the partition of unity property of scaling function eq (\ref{eq:partition_of_unity_1}).
\begin{eqnarray}
\int \delta(x)dx&=&1\nonumber\\
\sum_{m,n}\int \delta(x)s_{m}(x)s_{n}(x)dx&=&1\quad [\text{Using eq (\ref{eq:partition_of_unity_1})}]\nonumber\\
\label{eq:delta_scaling_inhomogeneous_equation}
\sum_{m,n} I_{ss,mn}&=&1
\end{eqnarray}
Here, the delta function is centred at the origin. So, the delta function and two scaling functions will overlap for $-4\leq m,n\leq -1$.
For all other values of $m$ and $n$ there will be no overlap among them, so the value of the integration will be $0$. We can solve the set of equations (\ref{eq:delta_scaling_homogeneous_equation}) and eq (\ref{eq:delta_scaling_inhomogeneous_equation}) to get all the possible values of integrals. See table (\ref{Tab:delta_and_scaling_function_overlap}).
\begin{table}[h]
\begin{tabular}{|c|c|}
\hline
Integrals & Values \\
\hline
$I_{ss,(-4)(-4)}$ & $0.0000179297$ \\

$I_{ss,(-4)(-3)}$ & $0.000403396$ \\

$I_{ss,(-4)(-2)}$ & $-0.00163377$\\

$I_{ss,(-4)(-1)}$ & $0.00544679$\\

$I_{ss,(-3)(-3)}$ & $0.00907591$\\

$I_{ss,(-3)(-2)}$ & $-0.0367577$\\

$I_{ss,(-3)(-1)}$ & $0.122546$\\

$I_{ss,(-2)(-2)}$ & $0.14887$\\

$I_{ss,(-2)(-1)}$ & $-0.496316$\\

$I_{ss,(-1)(-1)}$ & $1.65466$\\
\hline
\end{tabular}
\caption{\label{Tab:delta_and_scaling_function_overlap} The values of overlap integrals of product of delta function and two scaling functions.}
\end{table}

\bibliographystyle{unsrt}
\bibliography{References}
\end{document}